\documentclass[prl,twocolumn,longbibliography]{revtex4}

\usepackage[utf8]{inputenc}
\usepackage{bm}
\usepackage{mathtools}
\usepackage{bbold}
\usepackage{amsmath}
\usepackage{amssymb}
\usepackage{hyperref}
\usepackage[normalem]{ulem}
\usepackage{color}
\usepackage{float}
\usepackage[dvipsnames]{xcolor}

\begin{document}

\title{Scalable spin squeezing from critical slowing down in short-range interacting systems}
\author{Tommaso Roscilde$^1$, Filippo Caleca$^1$, {Adriano Angelone$^{2, 3}$,} and Fabio Mezzacapo$^1$} 
\affiliation{$^1$Univ Lyon, Ens de Lyon, CNRS, Laboratoire de Physique, F-69342 Lyon, France}
\affiliation{{$^2$Sorbonne Universit\'e, CNRS, Laboratoire de
Physique Th\'eorique de la Mati\`ere Condens\'ee, LPTMC,
F-75005 Paris, France}}
\affiliation{{$^3$eXact lab s.r.l., Via Francesco Crispi 56 - 34126 Trieste, Italy}}


\begin{abstract}
Long-range spin-spin interactions are known to generate non-equilibrium dynamics which can squeeze the collective spin of a quantum spin ensemble in a scalable manner, leading to states whose metrologically useful entanglement grows with system size.  Here we show theoretically that scalable squeezing can be produced in 2d U(1)-symmetric systems even by short-range interactions, {i.e. interactions that at equilibrium do not lead to long-range order at finite temperatures,} but rather to an extended, Berezhinski-Kosterlitz-Thouless (BKT) critical phase. If the initial state is a coherent spin state in the easy plane of interactions, whose {energy} corresponds to a thermal state in the critical BKT phase, the non-equilibrium dynamics exhibits critical slowing down, corresponding to a power-law decay of the collective magnetization in time. {This} slow decay protects scalable squeezing, whose scaling reveals in turn the decay exponent of the magnetization.  Our results open the path to realizing massive entangled states of potential metrological interest in many relevant platforms of quantum simulation and information processing -- such as Mott insulators of ultracold atoms, or superconducting circuits -- characterized by short-range interactions in planar geometries. 
\end{abstract}
\maketitle

\emph{Introduction.} Multipartite entanglement \cite{Horodeckietal2009,Guehne_2009,Pezze2018RMP, Frerotetal2023} is the central feature of some of the most extreme many-body states of qubit (or $S=1/2$ spin) ensembles -- such as total spin singlets, squeezed states or Schr\"odinger's cat states. Its controlled generation is an important perspective for quantum simulation and computing, as multipartite entanglement represents {both an} efficiently certifiable form of quantumness,  {and a} central resource for quantum metrology \cite{Pezze2018RMP}. A most time-efficient way of preparing multipartite entangled states is via a quantum quench, namely the evolution of an initial state $|\psi(0)\rangle$ { governed by} a time-independent many-body Hamiltonian ${\cal H}$, acting on all qubits in parallel. The time evolution of multipartite entanglement along the dynamics is then fundamentally dictated by the geometry of the couplings among qubits, i.e. by the dimensionality of the system as well as by the range of the interactions. 
In this respect, some of the most relevant platforms for quantum simulation and computing have a planar (i.e. 2d) geometry and short-range interactions, which we here define as interactions conforming with the Mermin-Wagner theorem \cite{MerminWagner, Bruno2001}, i.e. not leading to the appearance of long-range order at finite temperature in the equilibrium phase diagram of the system.  This is the case \emph{e.g.} of neutral atoms in quantum gas microscopes interacting via contact interactions \cite{GrossB2021}, and via Rydberg-dressed \cite{Guardado-Sanchezetal2021, Eckneretal2023} or van-der-Waals \cite{Bernien2017,Scholl2021,Ebadi2021} interactions; as well as of most superconducting circuit architectures \cite{Morvan2022,Kim2023}. Hence exploring the ability of two-dimensional short-range interacting systems to develop multipartite entanglement in their dynamics is of high experimental relevance. In particular{,} in this work we will focus on \emph{scalable} multipartite entanglement, involving a growing number of qubits/spins when the number of degrees of freedom increases; and on the scaling of the time necessary for the quench evolution to reach {optimal scalable squeezing}. In the case of short-range interacting systems, the search for scalability raises fundamental questions about the connection between the dynamical scaling of multipartite entanglement and the light-cone picture of correlation spreading imposed by causality  \cite{Chenetal2023}.   

Here, using a combination of time-dependent variational Monte Carlo \cite{Comparin2022PRA} and time-dependent rotor/spin-wave theory \cite{Roscildeetal2023} we show that two-dimensional arrays of qubits with short-range interactions can indeed develop scalable multipartite entanglement when evolved away from equilibrium, in the form of scalable spin squeezing. We demonstrate that the scalability of spin squeezing is fundamentally protected by the slow relaxation of magnetization in two-dimensional system, related to the existence of the Berezhinskii-Kosterlitz-Thouless critical phase \cite{Chaikinbook}. When relaxation occurs towards a critical phase, the magnetization displays critical slowing down, which continuously controls the scaling behavior of squeezing, up to the scaling observed in the one-axis-twisting model \cite{Kitagawa1993PRA,Ma2011PR} of all-to-all interacting qubits with $U(1)$ symmetry. Our results unveil a deep connection between thermodynamics and entanglement dynamics in many-body systems \cite{Blocketal2023}; and they have potential implications for the realization of squeezed states in many devices based on planar lattices of qubits, such as atomic clocks based on neutral atom arrays \cite{Eckneretal2023}.   

\emph{Spin squeezing from quench dynamics.} In this work we focus on a specific form of multipartite entanglement, corresponding to squeezing of the collective spin of a qubit ensemble. Defining the collective spin operator ${\bm J} = \sum_{i=1}^N {\bm S}_i$ of $N$ qubits, their state exhibits spin squeezing if the spin squeezing parameter
\begin{equation}
\xi_R^2 = \frac{N \min_{\perp} {\rm Var}(J_{\perp})}{|\langle {\bm J} \rangle|^2}
\end{equation}
is smaller than unity. Here the numerator contains the variance of a collective spin component ${\rm Var}(J_{\perp})$ minimized over all the directions perpendicular to the average spin orientation $\langle {\bm J} \rangle$. The spin squeezing parameter is a witness of multipartite entanglement as the condition $\xi_R^2 < 1/k$, with $k$ an integer, allows one to conclude that the state features at least $(k+1)$-partite entanglement \cite{Pezze2018RMP}, namely at least $k+1$ entangled spins whose state is not separable into smaller blocks. 

We consider a time evolution driven by the Hamiltonian ${\cal H}$, $|\psi(t)\rangle = e^{-i{\cal H} t} |\psi(0)\rangle$, which leads to squeezing, starting from a coherent spin state (CSS) $|\psi(0)\rangle = |{\rm CSS}\rangle = | \rightarrow_x \rangle^{\otimes^N}$ with $\langle \bm J \rangle = \langle J^x \rangle = N/2$ at $t=0$.  Along such an evolution the minimum variance $\min_{\perp} {\rm Var}(J_{\perp})$ generically reaches a non-zero minimum at a given time $t_{\rm min}$, because of the bounded nature of spin operators. A necessary condition for squeezing to be \emph{scalable}, namely stronger the bigger the system size, is that the minimum variance per spin scales to zero with $N$ at $t_{\rm min}$, i.e. $v_{\perp,\min} = (\min_{\perp} {\rm Var}(J_{\perp}) /N)_{t_{\rm min}} \sim N^{-\nu_0}$; and, in systems with local interactions, this is expected to happen at an optimal time  scaling with system size, $t_{\rm min} \sim N^{\mu}$, necessary for scalable entanglement to set in. The scaling of  $v_{\perp,\min}$ does not immediately translates into scalable spin squeezing, because one needs to consider as well the scaling of the magnetization per spin $m^x = \langle J^x \rangle/N$ at $t_{\rm min}$, $m^x_{\rm min} = m^x_{t_{\rm min}}$. For generic Hamiltonians ${\cal H}$ leading to thermalization of the evolved state at long times \cite{dalessio_quantum_2016}, the scaling of the magnetization is then strongly dependent on the thermodynamics of the system \cite{Blocketal2023, trifa2023scalable}. In this work we will be concerned with two-dimensional XXZ models, with Hamiltonian
  \begin{equation}
  {\cal H}_{\rm XXZ} = - \sum_{i<j} {\cal J}_{ij} \left ( S_i^x  S_j^x + S_i^y  S_j^y + \Delta S_i^z  S_j^z \right) 
  \label{e.Ham}
  \end{equation}
where the $i, j$ indices run over the $N = L\times L$ sites of a square lattice, and ${\cal J}_{ij}$ is the spin-spin coupling, to be specified later. In this work we will focus  on easy-plane anisotropies, namely $-1 \leq \Delta < 1$, and ferromagnetic couplings for the $x$ and $y$ spin components, namely ${\cal J}_{ij}>0$. 
Our main tool to analyze the quench dynamics governed by the above Hamiltonian is time-dependent variational Monte Carlo (tVMC) based on the pair-product wavefunction \cite{Comparinetal2022, PRB2019}, which has proven to be very accurate in reproducing the dynamics of the XXZ model with power-law interactions \cite{Comparin2022PRA, Comparinetal2022b, Roscildeetal2023}; and which {is} also accurate for the case of short-range interactions in 2d -- see Supplemental Material (SM) \cite{SM}, {where we  compare our results with exact diagonalization as well as with those of a semi-classical phase-space method} \cite{Perlin2020PRL,Youngetal2023,Muleadyetal2023}. 

\begin{figure}[ht!]
\begin{center}
\includegraphics[width=0.9\columnwidth]{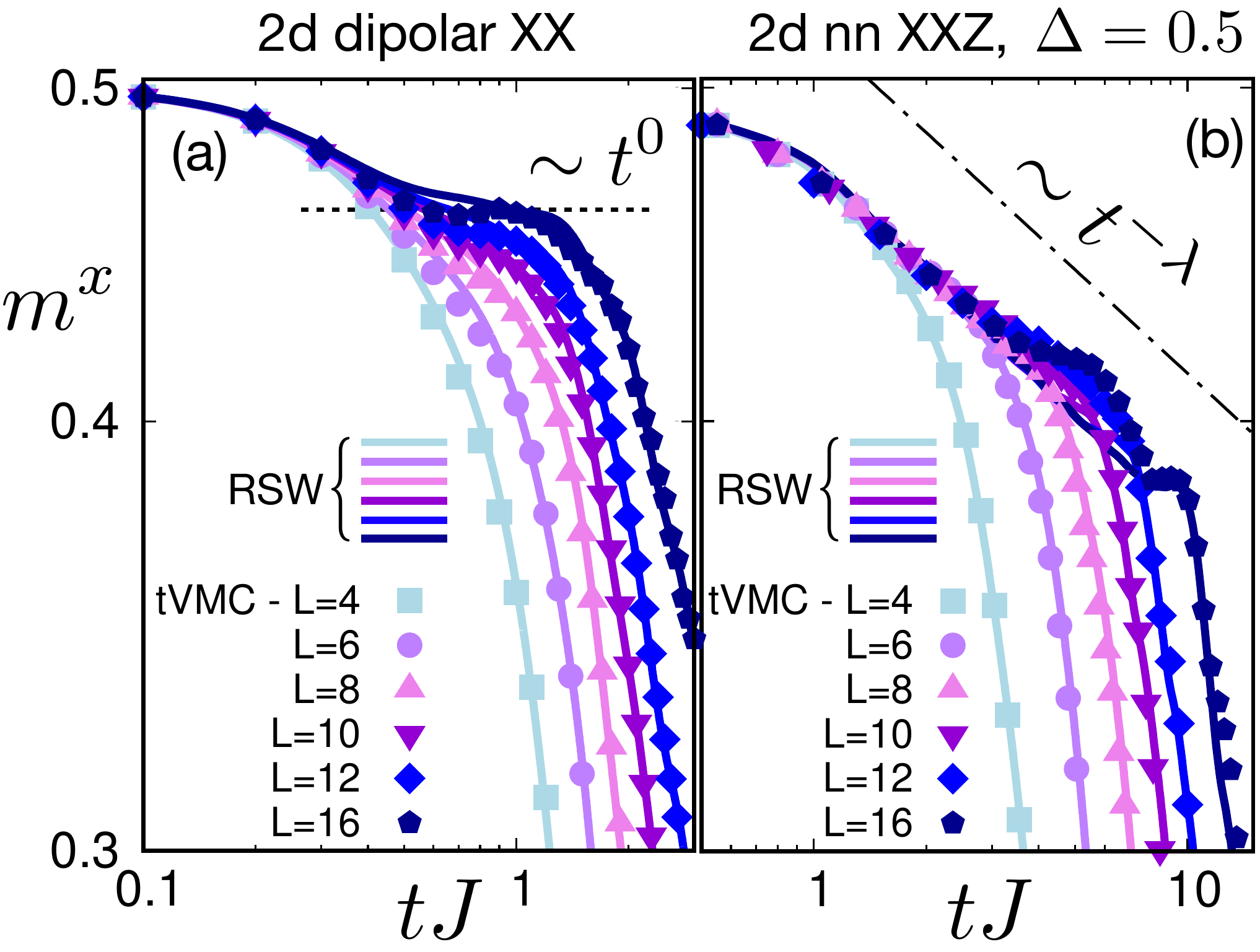}
\caption{Magnetization dynamics in a system whose thermalized state displays (a) long-range ordering or (b) critical BKT behavior. 
The tVMC and RSW data refer to (a) the 2d XX model with dipolar interactions \cite{Roscildeetal2023} and (b) the 2d XXZ with nearest-neighbor (nn) interactions. Here $\lambda = 0.1$, consistent with the scaling of optimal squeezing, see Fig.~\ref{f.2dXXZ}.}
\label{f.magn}
\end{center}
\end{figure}

\begin{figure*}[ht!]
\begin{center}
\includegraphics[width=0.9\textwidth]{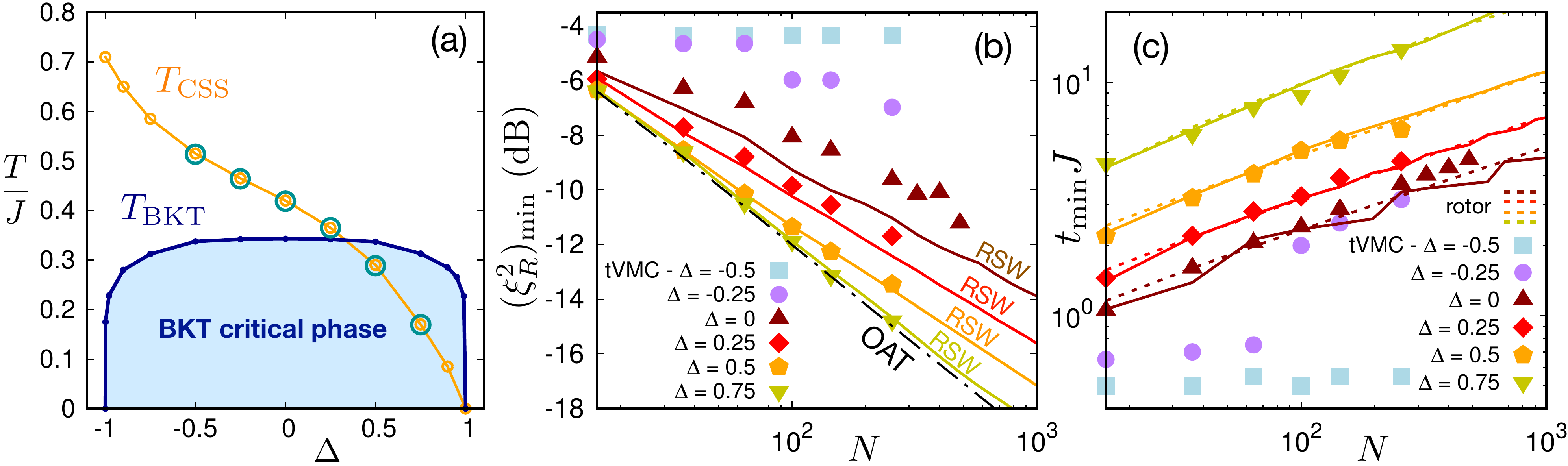}
\caption{Scalable spin squeezing in the nearest-neighbor 2d XXZ model. (a) Phase diagram from QMC -- the green circles correspond to the $\Delta$ values whose squeezing is analyzed in the other panels; (b) Scaling of optimal squeezing, comparing tVMC data with the predictions of RSW theory. The dot-dashed line indicates the OAT limit; (c) Scaling of the optimal squeezing time; solid lines are RSW predictions, and dashed lines (marked as 'rotor') are the predictions for the OAT model corresponding to the rotor degree of freedom.}
\label{f.2dXXZ}
\end{center}
\end{figure*}

\emph{Magnetization dynamics and thermodynamics.} 
As the two-dimensional XXZ model is not integrable, one can generally expect the non-equilibrium dynamics to relax towards a regime in which local quantities reflect those associated with thermal equilibrium at a temperature $T_{\rm CSS}$ corresponding to the energy of the initial state \cite{dalessio_quantum_2016}, namely $\langle {\rm CSS} | {\cal H} | {\rm CSS} \rangle =  {\rm Tr} \left ( {\cal H} e^{-{\cal H}/(k_B T_{\rm CSS}) } \right)/{\rm Tr} \left ( e^{-{\cal H}/(k_B T_{\rm CSS}) } \right) $.   The spatial structure of the couplings ${\cal J}_{ij}$ strongly affects the behavior of the system at the $T_{\rm CSS}$ temperature; and this in turn has a fundamental effect on the way in which the magnetization, initially imposed on the system with the choice of the CSS as initial state, decays during the quench evolution. For power-law decaying couplings,  ${\cal J}_{ij} = J/r_{ij}^\alpha$ with $\alpha<4$ \cite{Bruno2001}, the thermal phase diagram of the 2d system can feature ferromagnetic long-range order at low temperature. If $T_{\rm CSS}$ falls in the ordered phase of the system, then the dynamics is expected to feature a more persistent magnetization the bigger the system size -- a reflection of spontaneous symmetry breaking at $T_{\rm CSS}$ in the thermodynamic limit. This is the case e.g. of the dipolar XX ferromagnet ($\alpha=3$, $\Delta = 0$) depicted in Fig.~\ref{f.magn}(a). There the magnetization, {estimated} via tVMC, is seen to increase with system size at intermediate time, displaying a plateau that will last indefinitely in the thermodynamic limit -- this behavior was also observed recently in dipolar Rydberg atoms \cite{Bornetetal2023}. As a consequence, $m^x_{\rm min}$  scales as $m^x_{\rm min,\infty} - a N^{-\sigma} $, leading to a spin squeezing parameter which displays an even faster scaling than that of the minimum variance at moderate $N$ \cite{Comparinetal2022}, and crossing over to the asymptotic scaling $(\xi_R^2)_{\rm min} \sim N^{-\nu_0}$ for $N$ value such that $aN^{-\sigma} \ll {m^x_{\rm min,\infty}}$. This is also the case of the paradigmatic one-axis-twisting (OAT) Hamiltonian \cite{Kitagawa1993PRA}, corresponding to Eq.~\eqref{e.Ham} with \emph{e.g.} $\Delta = 0$ and ${\cal J}_{ij} = \frac{1}{I}$ (all-to-all couplings), for which $\nu_0 = 2/3$ \footnote{Even though strictly speaking the OAT Hamiltonian does not display thermalizing dynamics; and the dipolar XX Hamiltonian displays many features of the non-thermalizing dynamics of the OAT model, as shown by some of us in Refs.~\cite{Comparinetal2022, Roscildeetal2023}.}.

On the other hand, for short-range interactions, i.e. for $\alpha > 4$ -- which is the focus of this study -- Mermin-Wagner theorem \cite{Bruno2001} forbids spontaneous symmetry breaking at finite temperature, and the phase diagram {of the Hamiltonian in Eq.~\eqref{e.Ham}} features a paramagnetic phase at high temperature, and a critical, Berezhinski-Kosterlitz-Thouless (BKT) phase at low temperatures ($T < T_{\rm BKT}$), possessing quasi-long-range order. If $T_{\rm CSS}$ falls in the paramagnetic phase ($T_{\rm CSS} > T_{\rm BKT}$), then the magnetization is expected to decay to zero, because of the proliferation of topological defects (vortex/anti-vortex pairs). Generically, we can expect that $m^x = f(t/\tau)$ with a characteristic decay time $\tau$ which marks the vanishing of $m^x$, namely such that $f(1) = 0$. If $\tau$ is independent of the size \cite{SM}, then spin squeezing cannot scale indefinitely to lower values, since an increasing $t_{\rm min}$ will eventually hit the condition $t_{\rm min} = \tau$ at which $\xi_R^2$ diverges. 

 On the other hand, if $T < T_{\rm BKT}$, then the system is expected to thermalize towards a critical phase, with power-law-decaying correlations. This implies that two-point correlations will take more time to establish the larger the system size, inducing scaling in the thermalization dynamics. In the thermodynamics limit on{e} expects to observe \emph{critical slowing down}, namely a divergence of $\tau$, leaving behind a power-law relaxation of the magnetization $m^x \sim t^{-\lambda}$, with a continuously varying exponent $\lambda$ across the BKT phase. Such a behavior has been clearly observed in the stochastic, Monte Carlo dynamics of classical spins subject to local uncorrelated baths \cite{Ozekietal2003}; yet its observation in the unitary dynamics of a quantum many-body system is much more challenging, and not present in the literature, to the best of our knowledge. Fig.~\ref{f.magn}(b) clearly shows the onset of critical dynamics for a 2d XXZ model quenched in the BKT phase (short-range interactions and $\Delta = 0.5$, see below for further details). For every system size, the size-dependent drop in the magnetization following the power-law decay happens at a time scaling linearly with $L = N^{1/2}$ (see SM \cite{SM}): and it is associated with the correlations having spread all over the system, and marking the thermalization of local properties. 

Critical slowing down of the magnetization relaxation has an immediate consequence  on the scaling of optimal squeezing. At $t_{\min} \sim N^{\mu}$, we have that 
$m^x \sim t_{\rm min}^{-\lambda} \sim N^{-\lambda\mu}$. As a consequence, at $t_{\rm min}$ we have that $(\xi_R^2)_{\rm min} = v_{\rm min, \perp}/(m^x_{\rm min})^2 \sim N^{-\nu}$ with an exponent $\nu = \nu_0 - 2\lambda \mu$ which is corrected by the critical dynamics of the magnetization with respect to the case of long-range ordering at $T_{\rm CSS}$, for which $\lambda = 0$.  
This observation establishes a fundamental link between critical relaxation dynamics and multipartite entanglement dynamics, which is the main result of our work. 
The fact that the magnetization exhibits a significant time dependence at $t_{\rm min}$ shifts in principle the minimum of $\xi_R^2$  to earlier times with respect of $\min_{\perp} {\rm Var}(J^{\perp})$; yet, as we show in the SM \cite{SM}, this shift vanishes in the thermodynamic limit.

\emph{Rotor-spin-wave picture for the scaling of squeezing.} The above result does not provide any explicit prediction for the exponents $\nu$ and $\mu$ in the case of the short-range interacting XXZ model. In order to make such a prediction, we can resort to rotor/spin-wave (RSW) theory \cite{Roscildeetal2023, Roscildeetal2023b} recently proposed by some of us. Time-dependent RSW theory applies to any time-dependent state $|\psi(t)\rangle$ which possesses a large overlap with the Dicke manifold of states with maximum collective-spin length $\langle {\bm J}^2 \rangle = N/2(N/2+1)$. The theory describes the dynamics projected onto the Dicke manifold as that of an angular momentum variable ${\bm K}$ with $\langle {\bm K}^2 \rangle = N/2(N/2+1)$ governed by the OAT Hamiltonian ${\cal H}_R = \frac{(K^z)^2}{2I}$ with bare coupling constant $1/(2I) = \frac{1-\Delta}{N(N-1)} \sum_{i<j} {\cal J}_{ij}$ (which can be renormalized, see SM \cite{SM}). The leakage of the state out of the Dicke manifold is instead described in terms of linear, finite momentum spin waves, associated with free bosonic fields $b_{\bm k\neq 0}, b_{\bm k \neq 0}^\dagger$. The coupling between the zero-momentum rotor variable and the finite-momentum spin waves can be neglected as long as the density of bosons is sufficiently weak; and, within the same approximation scheme, all quantities of interest can be decomposed into a contribution from the rotor degree of freedom, and from the spin-wave ones. In particular RSW predicts that ${\rm Var}(J^\perp) \approx {\rm Var}(K^\perp)$, namely the variance of the transverse spin components obeys the OAT dynamics of the rotor \cite{Kitagawa1993PRA}, and therefore it exhibits a scaling minimum $v_{\rm min, \perp}$ with scaling exponent $\nu_0 = 2/3$ at a time $t_{\rm min}$ whose scaling exponent is $\mu = 1/3$. On the other hand, the magnetization has an additive structure, $m^x = \langle K^x \rangle/N -  n_{\rm SW}$ where $n_{\rm SW} = \sum_{\bm k\neq 0} \langle b_{\bm k}^\dagger b_{\bm k}\rangle/N$ is the spin-wave density. Hence, according to RSW theory, the scaling exponent for optimal squeezing in the presence of critical slowing down is (for large $N$) $\nu = (1-\lambda)\frac{2}{3}$, i.e. a renormalized version of the OAT scaling exponent; and hence scalable squeezing occurs, provided that $\lambda < 1$.  

In the following we shall test these predictions for scalable spin squeezing in the presence of critical slowing down for two fundamental models of two-dimensional interacting qubits in quantum simulators: the short-range XXZ model, and the XX model with van-der-Waals and Rydberg-dressed couplings. 

\emph{2d nearest-neighbor XXZ model.}  The first model that we discuss is the 2d XXZ model with nearest-neighbor interactions, namely ${\cal J}_{ij} = J$ only if $i$ and $j$ are neighboring sites on the square lattice -- modeling e.g. spin-spin interactions in Mott insulators of ultracold spinful atoms \cite{Jepsen2020,Zhangetal2023} or qubit-qubit interactions in superconducting circuits \cite{Morvan2022}. We reconstruct the phase diagram of the model, namely the $\Delta$ dependence of the two crucial temperatures $T_{\rm BKT}$ and $T_{\rm CSS}$, by making use of quantum Monte Carlo (QMC) based on the stochastic series expansion \cite{Syljuasen2002PRE}. As shown in Fig.~\ref{f.2dXXZ}(a), the CSS temperature goes below the BKT one for $\Delta \gtrsim 0.3$, so that critical slowing down in the magnetization dynamics is expected for this parameter range -- and indeed observed, as already shown in Fig.~\ref{f.magn}(b) for $\Delta = 0.5$. Our tVMC data for the optimal squeezing at $(\xi_R^2)_{\rm min}$, shown in Fig.~\ref{f.2dXXZ}(b), exhibit indeed scalable squeezing in the same parameter range, with a continuously varying scaling exponent, approaching that of the OAT dynamics when $\Delta \to 1$. The continuously varying scaling controlled by critical slowing down is well reproduced by RSW theory, which is found to quantitatively capture the critical slowing down dynamics in Fig.~\ref{f.magn}(b). In particular we find that $\lambda \approx 0.1$ for $\Delta = 0.5$ (see Fig.~\ref{f.magn}(b)); and $\lambda = 0.045$ for $\Delta = 0.75$. Correspondingly we observe  that $\nu = 0.62(1)$ for $\Delta = 0.5$ and $\nu = 0.728(16)$ for $\Delta = 0.75$, to be compared with $\nu = 0.66$ and $0.72$ predicted by the formula  $\nu = \nu_0 - 2\lambda \mu$ -- we took $\nu_0 = 0.73$ as the effective exponent for the OAT model in the size range considered here. 
  
  Moreover the optimal squeezing time $t_{\rm min}$  is also correctly captured by RSW theory, as seen in Fig.~\ref{f.2dXXZ}(c), and it is found to scale as in the OAT model, according to the RSW prediction that the transverse variance $\min_{\perp} {\rm Var}(J^\perp)$ obeys the OAT dynamics.   
On the other hand, scalable squeezing seems to be lost for sufficiently small $\Delta$ (see $\Delta = -0.5$ in Fig.~\ref{f.2dXXZ}(b)), and RSW is no longer predictive in this regime, since spin-wave excitations are highly populated and they are also accompanied by vortex-antivortex excitations in the dynamics. Interestingly, for intermediate values of $\Delta$, and in particular for the important case of the XX model, ($\Delta=0$), scalable squeezing is still observed -- as we can test with tVMC up to significantly large system sizes ($L=22$, or $N=484$). In this regime RSW still predicts correctly the scaling of the optimal time but not that of squeezing. This is due to the fact that the decoupling between the rotor and spin-wave degrees of freedom breaks down, as shown by the fact that the $v_{\perp,\min}$ no longer follows the physics of the OAT model (see SM \cite{SM}). We expect that scalable squeezing out of the BKT regime is a transient regime, due to the fact that the correlation length $\xi$ is extremely large at $T_{\rm CSS}$ (we estimate $\xi(T_{\rm CSS}) \approx 14$ from QMC), so that effectively the dynamics remains critical for the system sizes we considered.  This scalability of squeezing, albeit possibly transient, remains an interesting entanglement resource, as XX interactions can also be obtained from Ising ones (i.e. involving only one spin component) in the presence of a large transverse field \cite{Frankeetal2023}.

\begin{figure}[ht!]
\begin{center}
\includegraphics[width=\columnwidth]{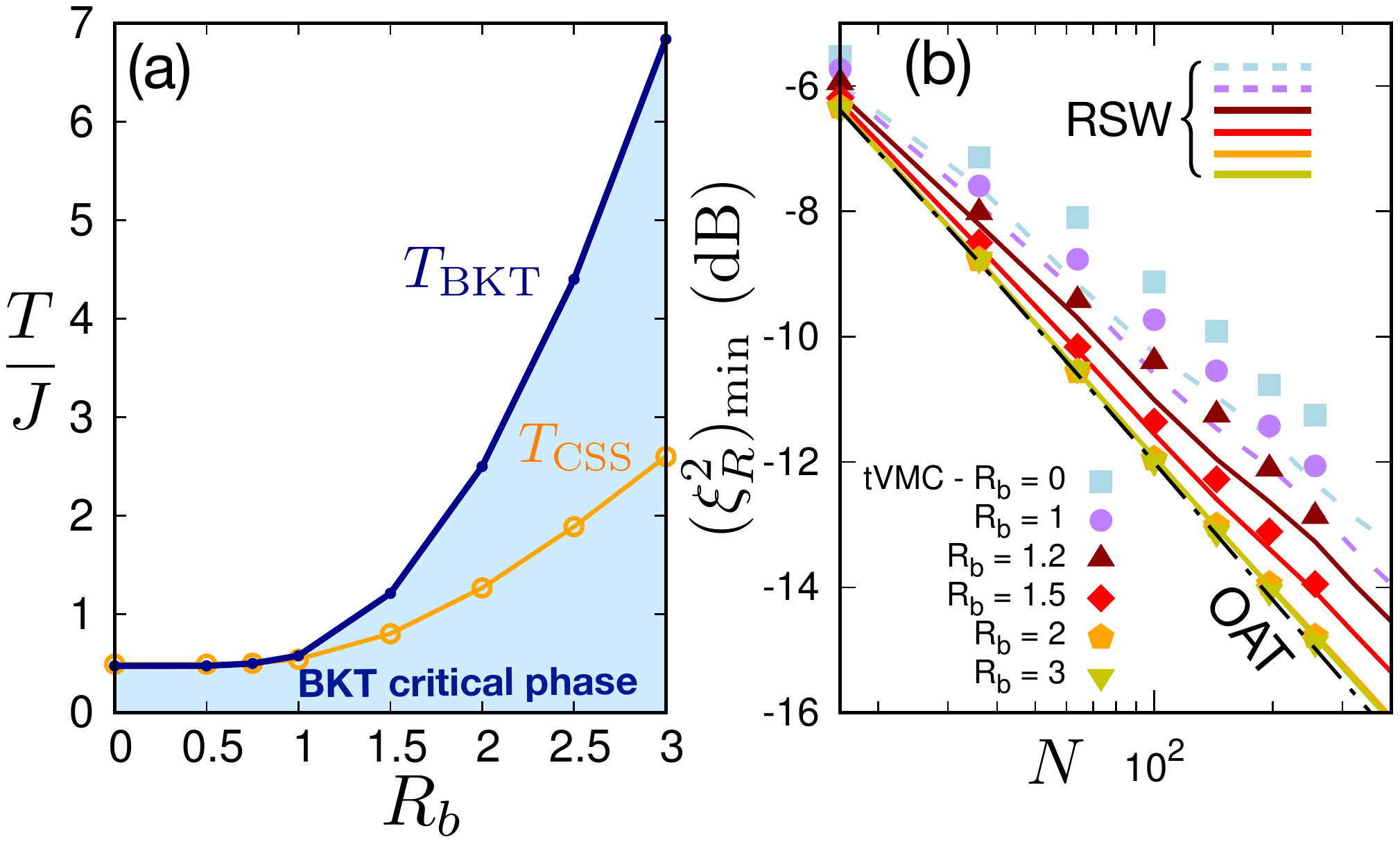}
\caption{Scalable spin squeezing in the 2d XX Rydberg-dressed model. (a) Equilibrium phase diagram; (b) Scaling of optimal squeezing. Significance of symbols as in Fig.~\ref{f.2dXXZ}(a-b).}
\label{f.Rydberg}
\end{center}
\end{figure}

\emph{2d XX model with Rydberg-dressed interactions.} 
We conclude this work by investigating the XX model ($\Delta=0$) with Rydberg-dressed interactions, ${\cal J}_{ij} = J (1+R_b^6)/(r_{ij}^6 + R_b^6)$ recently proposed for spin squeezing in Ref.~\cite{Youngetal2023} as realizable in arrays of Rydberg-atom clocks \cite{Eckneretal2023}. Here $R_b$ is the so-called blockade radius (in units of the lattice spacing), which is controlled by the ratio between the size of the Rydberg state and the inter-atomic distance. The limit $R_b=0$ gives the case of van der Waals interactions, while $R_b>0$ introduces a flat-top shape in the coupling at short distance. Fig.~\ref{f.Rydberg}(a) shows that this model possesses a critical BKT phase with a BKT temperature growing with $R_b$; and that $T_{\rm CSS}$ falls inside the BKT phase for $R_b\gtrsim 1$, {being} extremely close to the critical BKT temperature even for smaller $R_b$ {values}. The mechanism of critical slowing down is expected to lead to scalable spin squeezing for Rydberg-dressed XX interactions, at least for a sufficiently large $R_b$ value. In fact our tVMC results shown in Fig.~\ref{f.Rydberg}(b) appear to exhibit scalable squeezing for all the values of $R_b$ with a continuously varying scaling exponent $\nu$. For $R_b \gtrsim 1.5$, i.e. sufficiently deep in the BKT regime, RSW captures the scaling quantitatively and one can verify that $v_{\min,\perp}$ scales as in the OAT model (see SM \cite{SM}). On the other hand, the correct scaling is missed for smaller $R_b$ values as, similarly {to} what seen in the 2d XXZ model, RSW theory cannot capture the regime close to the BKT transition. 

\emph{Time optimality of 2d spin-squeezing dynamics.} Assuming that the interactions that we studied lead to a linear light-cone dynamics, then Lieb-Robinson bounds \cite{Chenetal2023} would impose that the entanglement depth cannot grow {faster}  than quadratically in time (in $d=2$), namely $\xi_R^{2} \geq \epsilon t^{-2}$ for some $\epsilon$. The OAT dynamics has the property that $(\xi_R^2)_{\rm min} \sim t_{\rm min}^{-2}$, which saturates this bound. This means that achieving a spin squeezing dynamics that approaches the OAT one with short-range interactions in 2d amounts to nearly reaching the fastest possible scaling allowed by Lieb-Robinson bounds.

\emph{Discussion and conclusions.} We have shown in this work that short-range interactions in 2d systems, i.e. interactions complying with the Mermin-Wagner theorem, can lead to scalable spin squeezing. This is due to an anomalously slow relaxation of the magnetization towards a critical BKT regime. We believe that ours is the first study of critical slowing down due to BKT physics in closed two-dimensional quantum systems, and we are able to show that such a phenomenon is not simply analogous to what observed in classical systems, but it is accompanied by the production of massive multipartite entanglement. Short-range interactions do not lead to scalable spin squeezing in 1d \cite{Comparin2022PRA, Blocketal2023}, so that $d=2$ represents the lower critical dimension for the appearance of scalable spin squeezing in short-range interacting systems. 
Our results indicate that many two-dimensional platforms for quantum simulation and computing possessing short-range interactions, including nearest-neighboring ones, can be envisioned as platforms for entanglement-assisted quantum metrology; and this has immediate potential consequences for devices such as optical-lattice atomic clocks \cite{SM, Ludlowetal2015,Bouganneetal2017, Franchietal2017, Campbelletal2017,Eckneretal2023}.

\begin{acknowledgements}
 \emph{Acknowledgements.} This work is supported by PEPR-Q (QubitAF project). We thank T. Comparin for his early contributions to this work. Useful discussions with L. Fallani, P. Holdsworth,  and A. M. Rey are gratefully acknowledged. All numerical simulations have been performed on the PSMN cluster at the ENS of Lyon.   
\end{acknowledgements}

\null
\newpage

\noindent
{\bf Supplemental Material} \\
\noindent
{\bf \emph{
Scalable spin squeezing from critical slowing down in short-range interacting systems}}

\begin{figure*}[ht!]
\begin{center}
\includegraphics[width=0.8\textwidth]{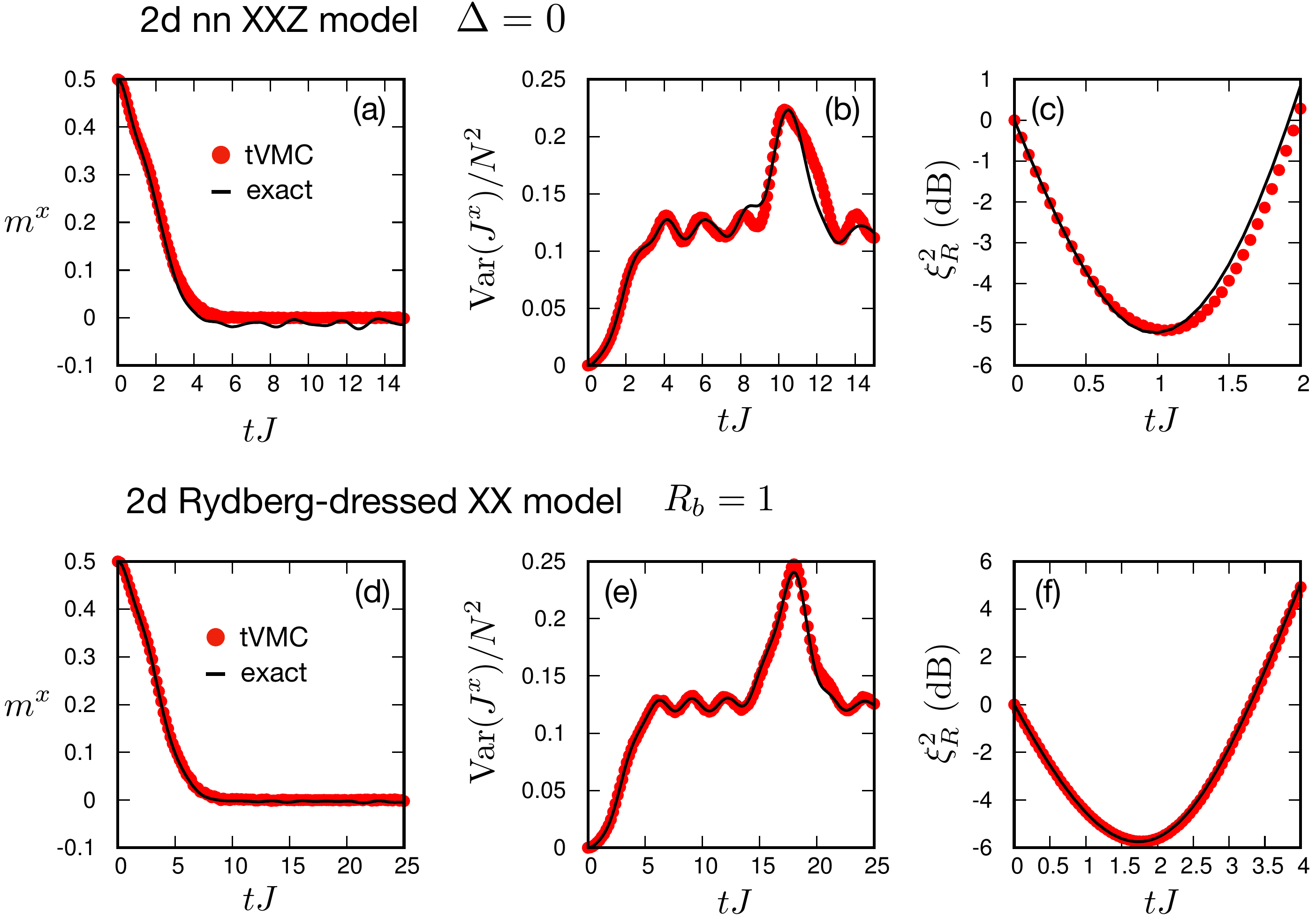}
\caption{{Comparison on a $N=4\times 4$ system between tVMC and exact diagonalization results for the two models of interest in this study and relevant values of the parameters $\Delta$ and $R_b$, respectively. (a,d) uniform magnetization; (b,e) variance of $J^x$; (c, f) squeezing parameter.}  }
\label{f.ED}
\end{center}
\end{figure*}

\begin{figure*}[ht!]
\begin{center}
\includegraphics[width=0.7\textwidth]{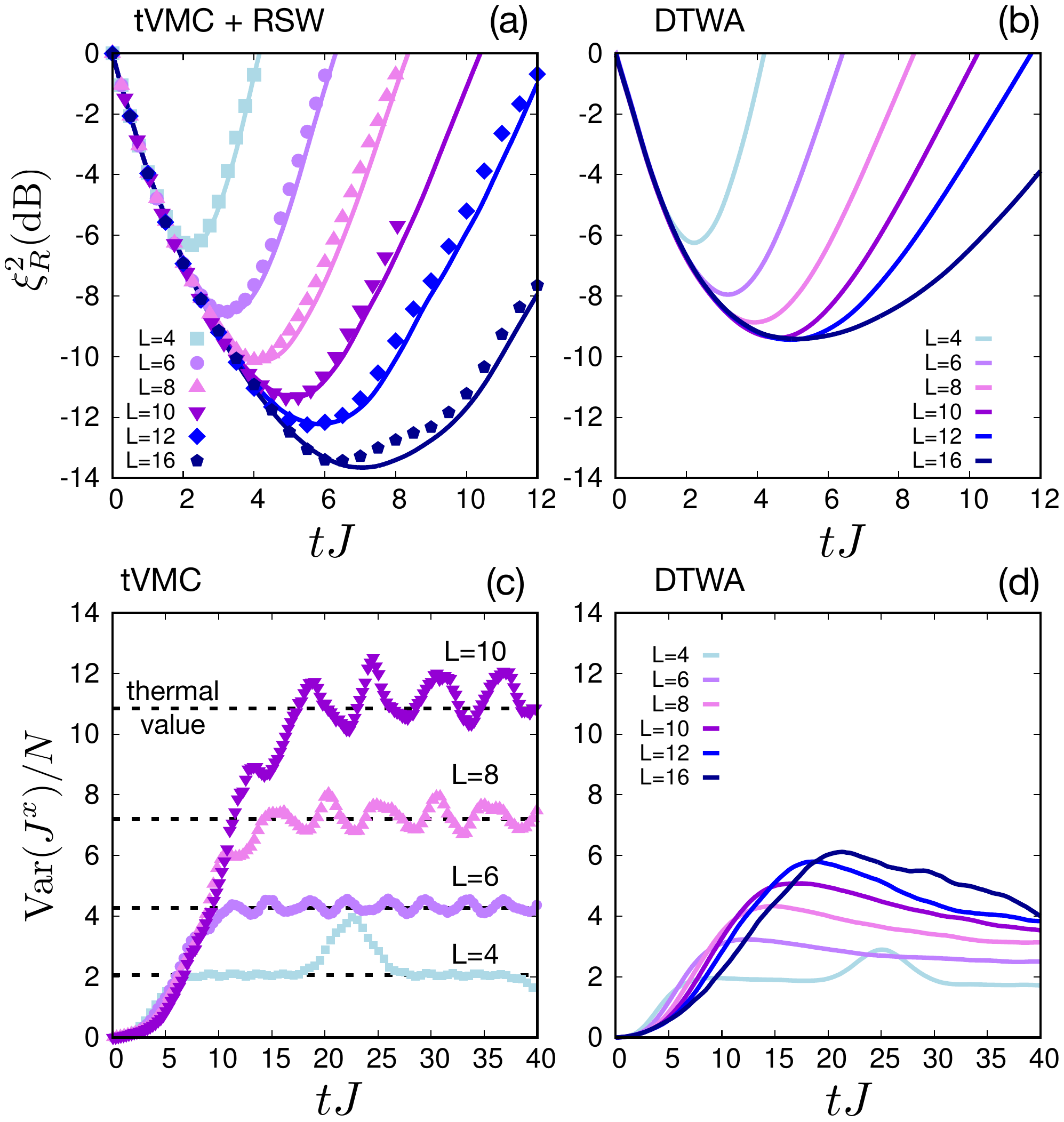}
\caption{Comparison of tVMC and RSW results with those of the discrete truncated Wigner approximation (DTWA) for the  2d XXZ model with nearest-neighbor interactions ($\Delta = 0.5$). 
(a-b) Spin squeezing dynamics from tVMC and RSW (solid lines) (a), and from DTWA (b). (c-d) Time evolution of ${\rm Var}(J^x)/N$ from (c) tVMC and (d) DTWA. In (c) the dashed lines indicate the thermal value at the temperature $T_{\rm CSS}$. }
\label{f.DTWA}
\end{center}
\end{figure*}

\section{Validation of the pair-product wavefunction, and comparison with the discrete truncated Wigner approximation}

{Fig.~\ref{f.ED} shows the comparison  between tVMC results based on the pair-product wavefunction \cite{Comparin2022PRA, PRB2019} and exact-diagonalization results on a system of $N = 4 \times 4$ sites,  for the two models investigated in this work and selected values of their parameters. We see that the pair-product wavefunction reproduces correctly not only the short-time squeezing dynamics, but also the full relaxation of the magnetization $m^x$ as well the development of correlations, captured by the variance of the collective spin component $J^x$ \cite{Comparinetal2022}.} 

Refs.~\cite{Perlin2020PRL, Muleadyetal2023} has investigated spin squeezing dynamics in the  2d XXZ model with variable-range interactions, by making use of the discrete truncated Wigner approximation (DTWA) \cite{Schachenmayer2015PRX}. The latter approximation appears to correctly reproduce the spin squeezing dynamics in two-dimensional models with power-law interactions \cite{Muleadyetal2023}. Yet, in the case of nearest-neighbor interactions \cite{Perlin2020PRL} it delivers different results compared to ours, most importantly it finds very weak or no scalable spin squeezing. We have implemented the DTWA approach following the the prescription of Ref.~\cite{Muleadyetal2023} for the choice of the initial Wigner function. In Fig.~\ref{f.DTWA}(a-b) we show the comparison between the tVMC and RSW results on the one hand, and the DTWA ones, for the case $\Delta = 0.5$. While tVMC and RSW are in very good agreement with each other and both predict scalable squeezing, as seen in the main text, we confirm that DTWA finds some scaling only for small sizes, but loss of scaling for larger ones.

 We justify this fundamental discrepancy between our results and DTWA in Fig.~\ref{f.DTWA}(c-d), showing the evolution of ${\rm Var}(J^x)/N$, which captures the onset of spin-spin correlations in the system. As discussed in the main text, from numerically exact quantum Monte Carlo calculations we expect thermalization to a BKT regime for $\Delta = 0.5$, namely to ${\rm Var}(J^x)/N \sim N^{1-\eta(T_{\rm CSS})}$ where $\eta(T_{\rm CSS}) \leq \frac{1}{4}$ is the exponent of the correlation function in the BKT phase \cite{Chaikinbook}, given that $T_{\rm CSS} < T_{\rm BKT}$. Fig.~\ref{f.DTWA}(c) shows that the tVMC results at long times capture very well the thermal values, which exhibit the super-extensive scaling characteristic of the BKT phase. On the other hand, the DTWA results only agree with the tVMC and thermal ones for very small systems ($L=4$), but for larger sizes they do not reproduce at all the correct thermal values. One can observe that DTWA fails to reproduce thermalization to a BKT phase, and rather predicts thermalization to a paramagnetic phase, with little to no scaling of ${\rm Var}(J^x)/N$. Hence in retrospect it is not surprising to see that DTWA is not capturing the presence of scalable spin squeezing, which we linked in this work to the critical relaxation towards a BKT regime.

\section{Decay of the magnetization in the paramagnetic phase and in the critical phase}

In the main text we have argued that, when thermalization of the dynamics takes place towards a paramagnetic phase, then one can expect a behavior of the magnetization in the form
$m^x = f(t/\tau)$ with a characteristic decay time $\tau$ marking the vanishing of $m^x$ ($f(1) = 0$), and that $\tau$ should be independent of system size. This assumption is based on the consideration that the local relaxation of the magnetization $m^x$ descends from the buildup of entanglement of each site with the other spins. Assuming that the establishment of two-point correlations is sufficient to lead to depolarization, then this mechanism is rather fast when thermalization occurs towards a short-range correlated state, since two-point correlations only need to spread up to a finite correlation length $\xi$ which is independent of system size. Under these circumstances, when $L \gg \xi$ the magnetization should exhibit relaxation to its vanishing equilibrium value within a size-independent time scale $\tau$. Upon approaching the phase transition to the BKT regime, on the other hand, the characteristic time scale for relaxation $\tau$ is expected to diverge, manifesting critical slowing down. When $\tau$ diverges, the only residual time scale is set by the system size, namely by the time that correlations take to cover the whole system. Such a time scale should scale linearly with $L$ if correlations spread ballistically in the system. 

\begin{figure}[ht!]
\begin{center}
\includegraphics[width=\columnwidth]{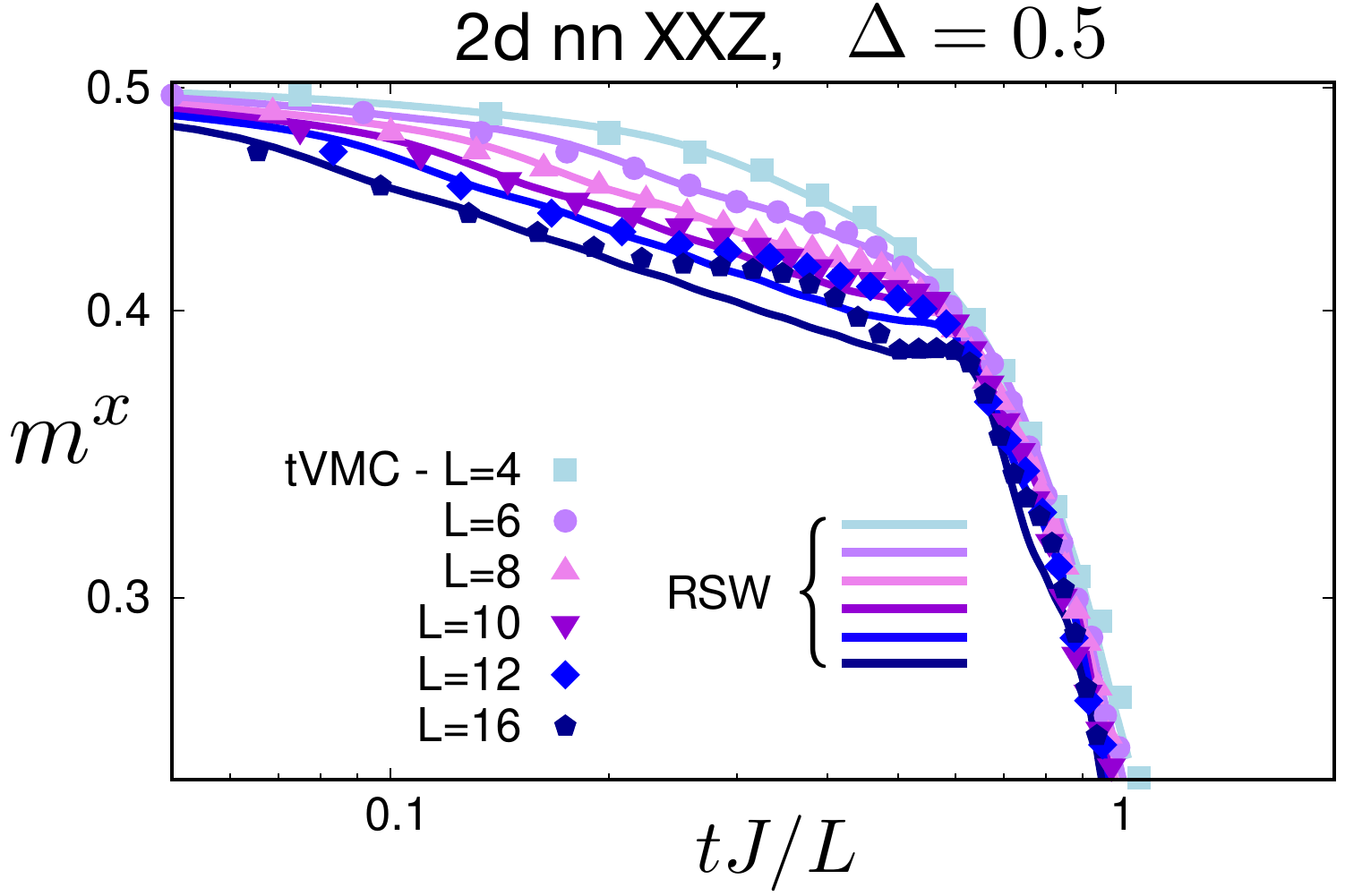}
\caption{Magnetization dynamics in the 2d XXZ model with nearest-neighbor interactions ($\Delta = 0.5$). Here we have rescaled the time by a factor given by the linear size $L$ with respect to Fig.~1 of the main text.}
\label{f.magn_toverL}
\end{center}
\end{figure}

Fig.~\ref{f.magn_toverL} shows the time dependence of the magnetization in the 2d XXZ model with nearest-neighbor interactions ($\Delta = 0.5$), with time rescaled by the linear size $L$ of the system. The sharp drop in magnetization, terminating the regime of power-law decaying magnetization $m^x \sim t^{-\lambda}$, occurs at a time growing linearly with linear size $L$, corroborating the argument offered above.   

\begin{figure*}[ht!]
\begin{center}
\includegraphics[width=0.7\textwidth]{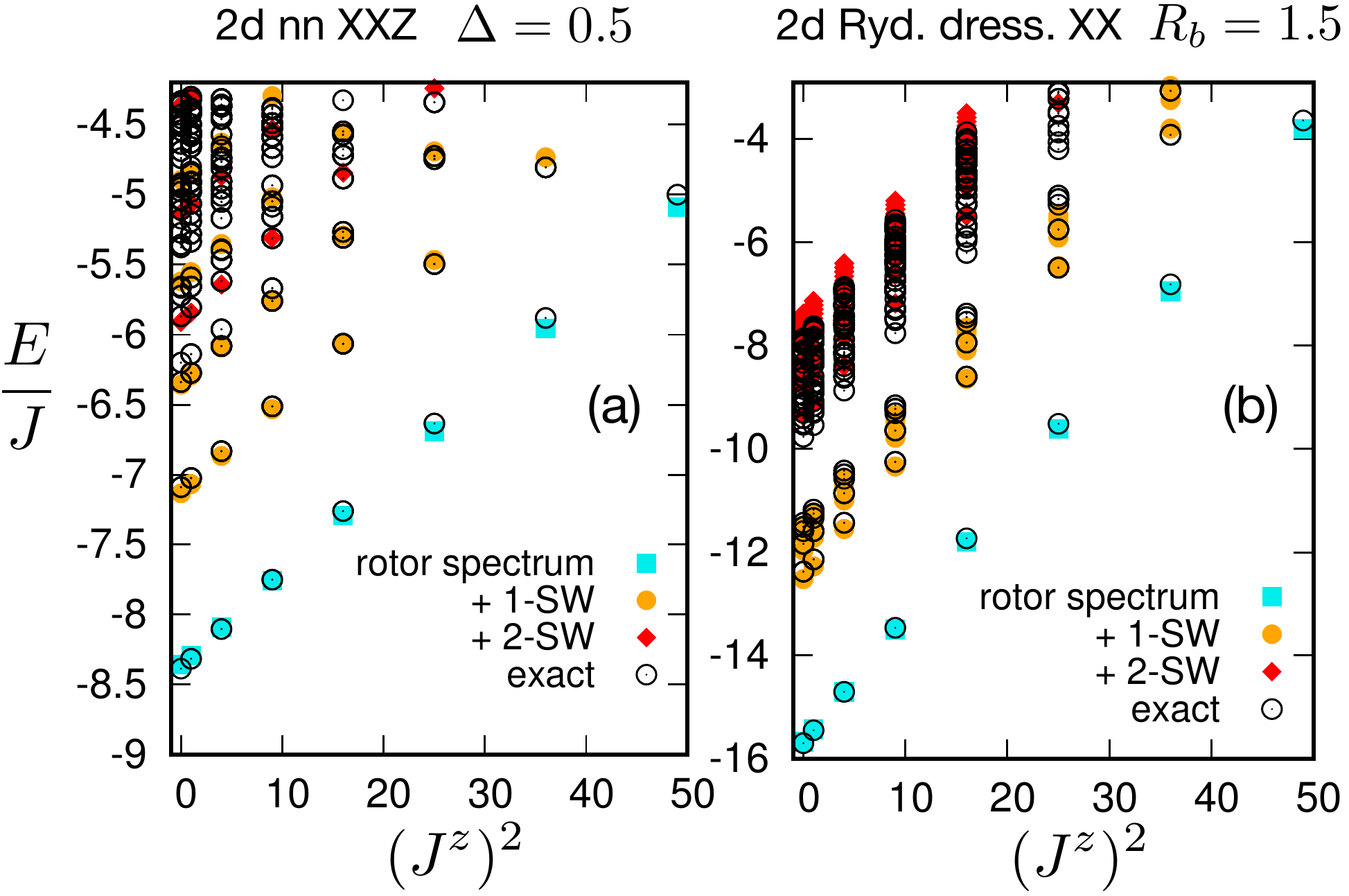}
\caption{Low-energy spectra for (a) the 2d XXZ model with nearest-neighbor interactions ($\Delta = 0.5$) and (b) the 2d XX model with Rydberg-dressed interactions ($R_b = 1.5$), calculated for a system with $N=4\times 4$ sites. On both panels the open circles indicate the results from exact diagonalization, while the closed symbols are the RSW predictions for the energy levels at various $J^z$ values in the presence of zero spin-wave (SW) excitations (i.e. the rotor spectrum), one SW excitation, and two (SW) excitations.}
\label{f.spectra}
\end{center}
\end{figure*}

\section{Optimal squeezing time in the presence of magnetization decay} 
\label{s.topt}

The argument which leads to the modified scaling of spin squeezing due to critical slowing down, namely $(\xi_R^2)_{\rm min} \sim N^{-\nu}$ with $\nu = \nu_0 - 2\lambda\mu$, assumes that the optimal squeezing is reached at the same time $t_{\rm min}$ at which the transverse variance reaches its minimum, $v_{\perp,\min}$. Given that $\xi_R^2(t) = v_{\perp}(t)/(m^x(t))^2$ this is clearly not obvious, because in principle the decay of the magnetization $m^x$ will shift the optimal squeezing time to earlier times with respect to $t_{\rm min}$. Yet we can show that in practice this shift is negligible in the thermodynamic limit. 

We can search for the time of optimal squeezing $t_{\rm opt}$ by imposing the vanishing of the logarithmic time derivative of $\log \xi_R^2$, namely
\begin{equation}
\frac{\partial \log \xi_R^2}{d\log t} \Big |_{t_{\rm opt}} = 0~. 
\end{equation}
Assuming $m^x\sim t^{-\lambda}$ this translates into the condition 
\begin{equation}
\frac{\partial \log v_{\perp}}{\partial \log t} \Big |_{t_{\rm opt}} = -2\lambda~. 
\end{equation}

For $t\approx t_{\rm min}$ we can assume that $v_{\perp}(t) \approx v_{\perp,\min} + a (t-t_{\min})^2$. Injecting into the above equation we immediately obtain that
\begin{equation}
t_{\rm opt} - t_{\min} = -\frac{2\lambda v_{\perp,\min}}{t_{\min}} + O(t_{\rm opt} - t_{\min})^2~.
\end{equation}
Given that $v_{\perp,\min}/t_{\min} \sim N^{-\nu_0 - \mu}$, the shift of $t_{\rm opt}$ with respect to $t_{\min}$ becomes negligible in the thermodynamic limit, justifying our treatment.

\section{Rotor/spin-wave theory: moment of inertia of the rotor from exact diagonalization} 

Rotor/spin-wave (RSW) theory \cite{Roscildeetal2023, Roscildeetal2023b} amounts to solving the dynamics of the lattice-spin model by mapping it to two independent degrees of freedom, namely 
${\cal H} \approx E_0 + {\cal H}_{\rm R} + {\cal H}_{\rm sw}$ where $E_0$ is the ground-state energy,  
\begin{equation}
{\cal H}_{\rm rot} = \frac{(K^z)^2}{2I}
\end{equation}
is the rotor Hamiltonian, and 
\begin{equation}
{\cal H}_{\rm sw} = \sum_{\bm k \neq 0} \omega_{\bm k} b_{\bm k}^\dagger b_{\bm k}
\end{equation}
the spin-wave one. The moment of inertia of the planar rotor is predicted as
\begin{equation}
\frac{1}{2I} = \frac{J_0 (1-\Delta)}{2(N-1)}~
\label{e.bare}
\end{equation}
while $\omega_{\bm k}$ is the dispersion relation of the spin waves, which for the $S=1/2$ Hamiltonian of interest in this work reads $\omega_{\bm k} = \sqrt{A_{\bm k}^2 - B_{\bm k}^2}$, where 
\begin{eqnarray}
A_{\bm k} & = & \frac{1}{2} \left [ J_0 - \frac{1}{2} J_{\bm k}(1+\Delta) \right ] \nonumber \\
B_{\bm k} & =&  - \frac{1}{2} J_{\bm k}(1-\Delta)~
\end{eqnarray}
and 
\begin{equation}
J_{\bm k} = \frac{1}{N} \sum_{ij} e^{i\bm k \cdot ({\bm r}_i - {\bm r}_j)} J_{ij} ~.
\end{equation}

The RSW prediction for the low-energy spectrum of the system is then of the form
\begin{equation}
E = E(K^z, n_{\bm k}) = E_0 + \frac{(K^z)^2}{2I} + \sum_{\bm k \neq 0} \omega_{\bm k} n_{\bm k}
\end{equation}
where $n_{\bm k}$ are the spin-wave populations. In particular the term in $(K^z)^2$ reconstructs the Anderson tower of states (ToS), corresponding to the ground-state energies of the system in the various $J^z$ sectors. With the identification $K^z \approx J^z$, one can compare quantitatively  spectra from exact diagonalization and the RSW spectra \cite{Roscildeetal2023b}. 
This is done in Fig.~\ref{f.spectra} for the 2d XXZ model with nearest-neighbor interactions and for the 2d XX model with Rydberg-dressed interactions. There we clearly observe that the low-energy spectrum for both models on $N=4\times 4$ lattices is well reproduced by RSW theory, which captures both the tower of states as the rotor spectrum, as well as the low-energy excitations in each $J^z$ sector, corresponding to spin waves. 

\begin{table}[ht!]
\begin{center}
\begin{tabular}{ c | c | c}
$\Delta$ &  $\frac{1}{2I}$ (RSW) & $\frac{1}{2I}$ (ToS) \\
\hline
$ -0.5$ & 0.2 & 0.236  \\
$ -0.25$ & 0.1666 & 0.188 \\
$ 0$ & 0.1333 & 0.145 \\
$ 0.25$ & 0.1& 0.107 \\
$ 0.5$ & 0.066 & 0.069
\end{tabular}
\caption{Inverse moment of inertia (in units of $J$) for the $4\times 4$ 2d XXZ model.}
\label{t.2dXXZ}
\end{center}
\end{table}
\begin{table}[ht!]
\begin{center}
\begin{tabular}{ c | c | c}
$R_b$ &  $\frac{1}{2I}$ (RSW) & $\frac{1}{2I}$ (ToS) \\
\hline
$ 0$ & 0.152 & 0.160  \\
$ 1$ & 0.1673 & 0.174 \\
$ 1.2$ & 0.190 & 0.196  \\
$ 1.5$ & 0.242 & 0.246  \\
$ 2$ & 0.337 & 0.339 \\
$ 3$ & 0.4603 & 0.461 \\
\end{tabular}
\caption{Inverse moment of inertia (in units of $J$) for the $4\times 4$ 2d XX model with Rydberg-dressed interactions.}
\label{t.2dRydberg}
\end{center}
\end{table}

At the same time one can notice that the RSW prediction for the ToS spectrum is slightly off with respect to the exact spectrum, and it can be simply corrected by a renormalization of the rotor moment of inertia, namely of the term $1/(2I)$. This also implies that an immediate improvement of RSW theory can be achieved by replacing the ``bare" moment of inertia, Eq.~\eqref{e.bare}, predicted by RSW theory with the one extracted from the slope of the ToS spectrum in the exact diagonalization data. As discussed in Ref.~\cite{Roscildeetal2023b}, this correction is expected to come from normal ordering of all terms coupling zero-momentum and finite-momentum bosons, which are altogether neglected within the standard RSW approach. Tables \ref{t.2dXXZ} and \ref{t.2dRydberg} report the predicted moment of inertia from Eq.~\eqref{e.bare} with the one extracted from the exact results for the ToS spectrum. We observe that the renormalization of the moment of inertia is more significant for small $\Delta$ in the 2d XXZ model and for small $R_b$ in the 2d XX model with Rydberg-dressed interactions; but it tends to disappear as $\Delta$  and $R_b$ increase. This suggest that the RSW theory predictions become increasingly accurate for $\Delta \to 1^- $  (in the XXZ model) and for increasing $R_b$ (in the Rydberg-dressed XX model). 

The improvement of RSW theory based on the renormalization of the moment of inertia is nonetheless limited to system sizes for which exact diagonalization is available. Yet, knowing the moment of inertia for a given system size $N$, we can reconstruct the one for a different system size by using the simple scaling formula (valid for the ``bare" moment of inertia, Eq.~\eqref{e.bare}) \cite{Comparinetal2022}
\begin{equation}
\frac{1}{2I_{N'}} = \frac{N'-1}{J^{(N')}_0} \frac{1}{2I_{N}}~.
\end{equation}
All the time-dependent RSW data shown in this work have been obtained by using effective moments of inertia for the rotor variable rescaled according to the above prescription, starting from the exact-diagonalization data for systems with $N=4\times 4$.

\begin{figure*}[ht!]
\begin{center}
\includegraphics[width=0.8\textwidth]{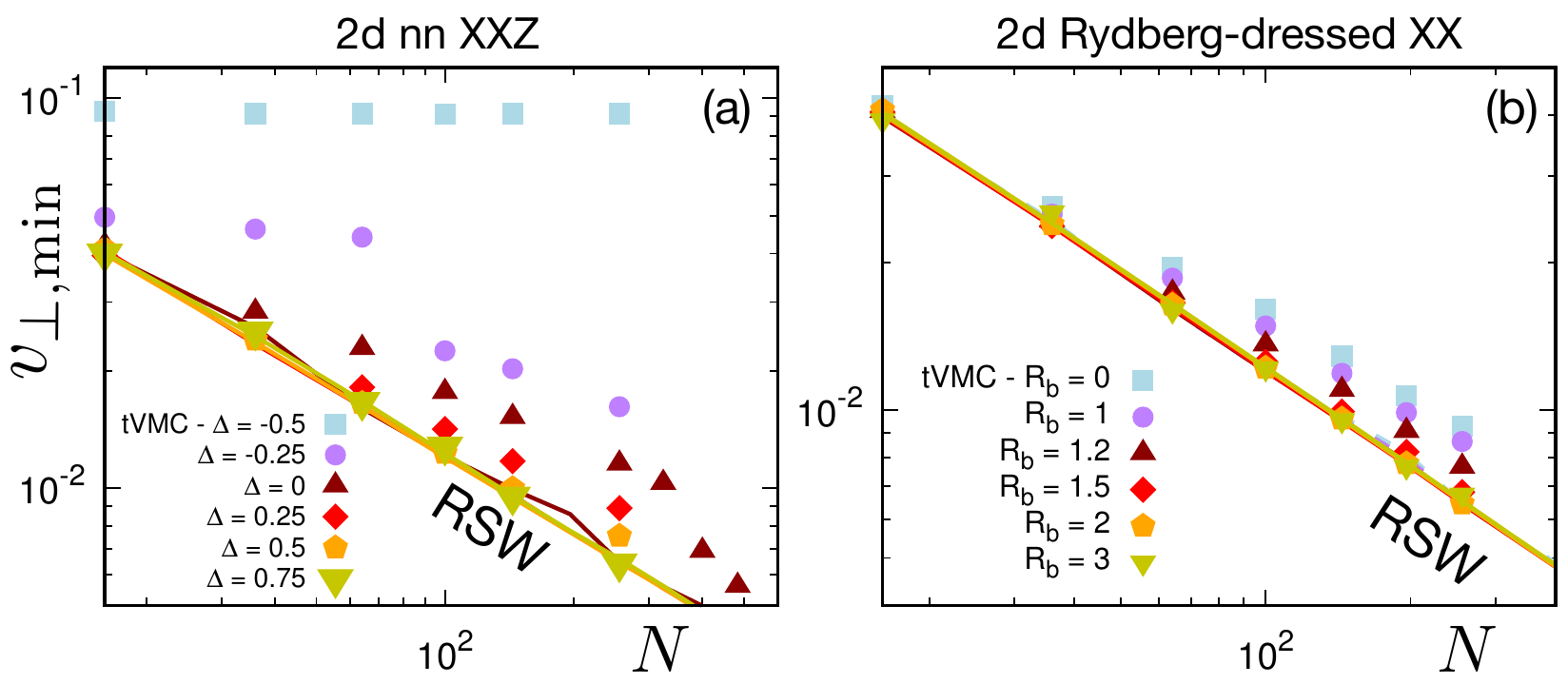}
\caption{Scaling of the transverse variance at the optimal squeezing time for the two models of interest in this work. In both panels solid lines show the predictions from RSW theory.}
\label{f.varminscaling}
\end{center}
\end{figure*}

\section{Scaling of the minimum variance in the 2d XXZ model and in the Rydberg-dressed XX model}

Fig.~\ref{f.varminscaling} shows the scaling of the minimal transverse variance $v_{\perp,\min}$ for the two models of interest in this work -- the value of the minimal variance is evaluated at the optimal time for squeezing $t_{\rm opt}$, see discussion in Sec.~\ref{s.topt}. RSW theory predicts that $v_{\perp}$ follows the OAT dynamics of a planar rotor, so that its value at the optimal time should be essentially model-independent. In Fig.~\ref{f.varminscaling} we can see that this becomes indeed the case for the 2d XXZ model with nearest-neighbor interactions as well as the 2d XX model with Rydberg-dressed interactions for model parameters such that the dynamics exhibits critical slowing down, thermalizing to a critical BKT phase -- e.g. for $\Delta = 0.5$ and $0.75$ in the 2d XXZ model, and for $Rb \gtrsim 1.5$ for the 2d Rydberg-dressed XX model. These results further validate the fundamental mechanism identified by RSW theory as being the source of  scalability for spin squeezing in these systems. 

\section{Importance of 2d geometries for optical atomic clocks}

The observation of scalable spin squeezing for two-dimensional lattice spin models with short-range interactions is particularly relevant for  \emph{optical} atomic clocks \cite{Ludlowetal2015,Bouganneetal2017, Franchietal2017, Campbelletal2017,Eckneretal2023}. 
On the one hand, the native interactions between spins (related to the clock transition) for cold atoms in optical lattices, or trapped in optical tweezer arrays, are nearest-neighbor ones or Rydberg dressed ones, as studied in the present work. 

Moreover, in order to produce useful entanglement in these devices, 2d geometries of interactions are preferrable whenever the clock-frequency wavelength is comparable to the interatomic spacing. This is due to the fact that a uniform CSS, initializing the squeezing dynamics described in this work, can only be produced by an optical Rabi pulse along two spatial dimensions (perpendicular to the clock laser). Along the propagation direction of the laser, instead, the orientation of the CSS varies, due to the spatial phase of the Rabi frequency, corresponding to a spiraling pattern. 
Such a pattern may not have the low energy required for the collective spin polarization to relax slowly, so that scalable spin squeezing is no longer protected.

\bibliography{squeezing.bib}

\begin{thebibliography}{42}
\bibitem[{SM()}]{SM}
\bibinfo{note}{See {S}upplemental {M}aterial ({SM}) for details about: 1)
  validation of the pair-product wavefunction in tVMC and comparison with the
  discrete truncated Wigner approximation; 2) discussion of the temporal decay
  of the magnetization; 3) the difference between the optimal squeezing time
  and the minimal variance time; 4) RSW theory and the extraction of the moment
  of inertia; 5) the scaling of the minimum variance; 6) the potential
  relevance of our results for optical atomic clocks.}
\end{thebibliography}

\end{document}